\documentclass[a4paper]{article}
\usepackage{eurosym}
\usepackage{amsfonts}
\usepackage{amssymb}
\usepackage{amsmath}
\usepackage{harvard}
\usepackage{cite}
\usepackage{float}
\usepackage{graphicx}
\usepackage{color}

\newtheorem{example}{Example}[section]

\newtheorem{remark}{Remark}[section]

\renewcommand{\cite}{\citeasnoun}

\begin{document}

\title{Decision Theory with a Hilbert Space as Possibility Space \thanks{%
We would like to thank the Marsilius Kolleg at the Universit\"{a}t
Heidelberg. Without the stimulating discussions with researchers from all
disciplines of the humanities and sciencies at this forum the research in
this paper would not have been possible. }}
\author{J\"{u}rgen Eichberger \\
Alfred Weber Institut, Universit\"{a}t Heidelberg \and Hans J\"{u}rgen
Pirner \\
Institut f\"{u}r Theoretische Physik, Universit\"{a}t Heidelberg}
\date{\today }
\maketitle

\begin{abstract}
In this paper, we propose an interpretation of the Hilbert space method used
in quantum theory in the context of decision making under uncertainty. For a
clear comparison we will stay as close as possible to the framework of SEU
suggested by \cite{Savage-1954}. We will use the \cite{Ellsberg-1961}
paradox to illustrate the potential of our approach to deal with well-known
paradoxa of decision theory.
\end{abstract}

\strut

\textbf{Keywords: }Decision theory,\textbf{\ }uncertainty, Ellsberg paradox,
quantum theory, Hilbert space, possibility space

\strut

\textbf{JEL Classification codes: }C00, C44, D03, D81

\strut

\section{Introduction}

\emph{Subjective Expected Utility (SEU)}\ theory (or \emph{Bayesian decision
theory}) in the spirit of \cite{Savage-1954} has become the almost
uncontested paradigm for decision theory in Economics and Statistics.

In Economics a \emph{decision problem} is described by a set of \emph{states
of the world} $s\in S$, a set of \emph{outcomes} (or \emph{consequences}) $%
\omega \in \Omega $ and a set of \emph{actions }$a\in A$. States in $S$ are
assumed to be mutually exclusive and to be observed perfectly after
realizations, e.g., for a set of states $S:=\{s_{1},s_{2}\}$ state $s_{1}$
could mean "it rains"' and state $s_{2}$ "`it does not rain". Outcomes are
the ultimate objects which a decision maker values, e.g., sums of money,
consumption bundles, work conditions, etc. In this sense, these sets contain
semantic information. \newline
In its most general form, actions $a,$ or acts as \cite{Savage-1954} calls
them, are mappings from states to consequences, $a:S\rightarrow \Omega .$
Hence, the set of potential actions is the set of all such mappings: $%
A:=\{a|\ a:S\rightarrow \Omega \}.$ A decision maker will choose an action $%
a $ $\in A$ before knowing which state $s\in S$ will be realized. Once a
state $s$ is realized (revealed) the decision maker who chooses action $a$
obtains outcome $a(s).$

According to the behavioristic approach of economics, one assumes that only
choices of actions can be observed. Hence, \cite{Savage-1954} assumes that
every decision maker's preferences over these actions can be characterized
by a complete \emph{preference order }$\succeq $ on $A.$ A decision maker is
assumed to choose $a$ rather than $b$ if and only if $a$ \emph{"is weakly
preferred to"} $b,\ $ $a\succeq b.$

In economic applications, working with objective functions that contain
parameters which one can give semantic interpretations simplifies the
analysis. Therefore economists have searched for conditions on the
preference order $\succeq ,$ axioms as \cite{Savage-1954} calls them,\
characterizing an objective function $V:A\rightarrow \mathbb{R}$ which
represents preference in the sense that $a\succeq b$ if and only if $%
V(a)\geq V(b).$

In a seminal contribution, \cite{Savage-1954} provides eight axioms on
preferences which guaranteed the existence of (i)\ a utility function $%
u:\Omega \rightarrow \mathbb{R}$ and (ii) a probability distribution $p$ on $%
S$ such that the \emph{S}ubjective \emph{E}xpected \emph{U}tility (\emph{SEU}%
) functional $V(a):=\int_{S}u(a(s)\ dp(s)$ represents this preference order $%
\succeq $. Since the probability distribution $p$ on $S$ is derived from the
subjective preferences of the decision maker over action in $A$ it is called
a \emph{subjective probability distribution}.

>From a normative point of view, the SEU model has become widely accepted as
the \emph{"rational" }way to deal with decision making under uncertainty. As
a descriptive model of behavior, however, the theory has been less
successful. Laboratory experiments have confronted decision theory with
behavior contradicting the assumption that subjects' beliefs can be modeled
by a subjective probability distribution and that their valuations of random
outcomes can be represented by expected utility. Psychological factors have
been put forward to explain this phenomenon (e.g., 
\citename{Kahneman-Tversky-1979} \citeyear*{Kahneman-Tversky-1979}%
). \cite{Ellsberg-1961} shows that people systematically prefer choices for
which information about probabilities is available to uncertain risks. Such
behavior violates the basic rules of probability theory.

In response to the Ellsberg paradox and many other discrepancies between
behavior observed in laboratory experiments and behavior suggested by SEU
theory, \cite{Schmeidler-1989} and \cite{Gilboa-Schmeidler-1989} started off
an extensive literature which tries to accommodate these paradoxes by
relaxing the axioms, mostly the independence axiom, proposed by \cite%
{Savage-1954}. \cite{Machina-Siniscalchi-2014} provide an excellent survey
of this literature. All these approaches remain, however, firmly rooted in
the Savage framework of states, acts, and consequences.

In this paper, we propose an interpretation of the Hilbert space method used
in quantum theory in the context of decision making under uncertainty. This
new approach departs from the Savage paradigm of states as "a description of
the world, leaving no relevant aspect undescribed" ( 
\citename{Savage-1954} \citeyear*{Savage-1954}%
, p. 9). In quantum mechanics the state of a system cannot be verified
directly in contrast to measurements which can be observed. For the sake of
a clear exposition, we will stay in our exposition as close as possible to
the framework of SEU suggested by \cite{Savage-1954}. We will use the
two-color urn version of the \cite{Ellsberg-1961} paradox to illustrate the
potential of our approach to deal with well-known paradoxes of decision
theory.

This article illustrates how one can implement an abstract Hilbert space as
possibility space in decision theory. We propose mathematical methods common
in quantum physics. The difficulties applying quantum mechanics to decision
theory come from the distinct terminologies used in both disciplines. Hence,
we will pay special attention to concepts in both fields which share names
but not meanings. In order to avoid misunderstandings it is necessary to
develop a dictionary which relates concepts from decision theory and quantum
mechanics. Quantum mechanics uses a mathematical formalism that quantifies
possibilities in the form of wave functions. Wave functions are elements of
a Hilbert space and represent probability amplitudes. The Hilbert space does
not provide probability distributions directly. We will argue in this paper
that mapping concepts from quantum mechanics to the theory of decision
making under uncertainty promises a framework in which one can deal with the
interrelation between subjective beliefs and objective information about
events in a non-Bayesian way.

Applying quantum mechanic methods to describe decision making behavior of
humans does not mean that "quantum physics determines human decisions".
Hence, it would be misleading to speak of "quantum decision theory". The
fact that electromagnetic waves and water waves can be described by the same
mathematical methods does not mean that electromagnetic waves consist of
water.

Though we hope that our proposed interpretation opens new ways for the
analysis of decision making under uncertainty, our focus in this paper is on
a suggested interpretation of the quantum mechanics framework in the context
of decision theory. Hence, we will not exploit the full power of these
methods. It appears, however, straightforward to extend our approach to more
general state spaces. Yet, such generalizations will be left for further
research.

\subsection{Literature}

There is a small literature which tries to apply methods of quantum
mechanics to decision making in Economics. All these papers use the same
mathematical framework of a Hilbert space of wave functions. The main
differences lie in the interpretation of the objects of the theory. In
particular, one needs to give the "wave functions" and their entanglement
and superpositions an interpretation in an decision context. Moreover,
quantum mechanics has no notion of choice over actions or acts. Hence,
differences concern mainly these aspects of the theory.

In a series of papers\footnote{\cite{Aerts-Sozzo-2011}, \cite%
{Aerts-Sozzo-2012}, \cite{Aerts-Sozzo-2012a}, \cite{Aerts-Sozzo-Tapia-2012}, 
\cite{Aerts-Sozzo-Tapia-2013}.}, Diederik Aerts, Sandro Sozzo and Jocelyn
Tapia propose a framework for modelling decision making under ambiguity by
quantum mechanic methods. \cite{Aerts-Sozzo-Tapia-2012} applies a version of
this approach to the three-color urn version of the Ellsberg paradox.
Similar to the approach suggested in this paper, these authors interpret the
elements of a complex Hilbert space as determinants of the probabilities and
actions as projectors from the basic elements of the Hilbert space to
outcomes or payoffs. In contrast to this paper however, they obtain
entanglement by projecting the pre-probabilities of the extreme compositions
of the urn.

In a second group of papers\footnote{\cite{Yukalov-Sornette-2010}, \cite%
{Yukalov-Sornette-2011}, \cite{Yukalov-Sornette-2012}.}, Vyacheslav Yukalov
and Didier Sornette model decision making under uncertainty also by a
Hilbert space of wave functions. \cite{Yukalov-Sornette-2011} apply their
model to several paradoxes of decision theory, in particular to an analysis
of the disjunction and the conjunction effect. Their interpretation of wave
functions as "intentions" or "intended actions" is quite different from the
interpretation suggested in this paper.

\section{Individual decision making and the mathematics of quantum
mechanics: a suggested interpretation}

In this section, we will describe economic decision in terms of the
mathematical concepts provided by quantum mechanics. The most fundamental
difference to Savage`s SEU approach sketched in the introduction is the fact
that states of the world are no longer observable after they are realized.
In quantum mechanics, "states of the world" in the Savage sense are
orthonormal vectors of a general Hilbert space. General elements of the
Hilbert space, which in physics are wave functions, represent states of the
world, because they encode all kind of information about the decision
context, including probability information about the orthonormal basis
states. Most importantly, these states will never be revealed or observed
directly. What can be observed however, in a general state are the
probabilities of the basis states. These probability distributions over
basis states can be obtained by projections of the general state onto the
basis states. As we will show in this section, actions will also contain
projectors on the basis states.

For simplicity, we choose as starting point a finite dimensional Hilbert
space $H:=\mathbb{C}^{|\Omega |}$ spanned by a complete orthonormal set of
basis states $\Psi _{\omega }:=|\omega >\in H$. In the following we will use
interchangeably the symbols $\Psi $ and the Dirac notation $|\Psi >$ for the
elements of the Hilbert space. General states of the world are linear
combinations of these basis states $|\beta >=\sum_{\omega =1}^{\Omega }\beta
_{\omega }\Psi _{\omega }\in H$. Each state contains several complex
amplitudes. Projectors of the composite state on basis states $P_{\omega
}:=|\omega ><\omega |$ give the probability of finding the basis state $%
\omega $ in the complex state $|\beta >$. Note, however, that the complex
coefficients of a state $|\beta >$ contain more information than
probabilities over outcomes.

An action $\alpha $ is a mapping $A_{\alpha }:=\underset{\omega }{\sum }%
\alpha _{\omega }P_{\omega }$ from composite states to outcome-specific
utility payoffs, i.e., a weighted sum of projection operators which map a
composite state $|\beta >$ into another state of the Hilbert space $\underset%
{\omega }{\sum }\alpha _{\omega }P_{\omega }|\beta >$. The average utility $%
U(\alpha ,\beta ) $ from an action operator $A_{\alpha }$ in a state $|\beta
>\ \in H$ is obtained by evaluating the expectation value of the action in a
composite state. In this sense, the state of the world $|\beta >$ allows to
identify a probability distribution over outcomes in $\Omega $. The expected
utility of action $\alpha $ in the state $|\beta >$ is: 
\begin{equation}
U(\alpha ,\beta ):=<\beta |A_{\alpha }|\beta >=\underset{\omega }{\sum }%
u(\alpha ,\omega )|\beta _{\omega }|^{2}.
\end{equation}

For a basis state $|\omega >,$ the action $\alpha $ reveals the utility
obtained from the action if the outcome $\omega $ would occur with
certainty:\\
$u(\alpha ,\omega ):=U(\alpha ,\omega )=<\omega|A_{\alpha
}|\omega >=\alpha _{\omega }.$

In any state $\beta $, decision makers are assumed to choose among a set of
actions $\mathcal{A}$ according to their expected utility $\arg \max
\{U(\alpha ,\beta )|\ \alpha \in \mathcal{A}\}.$

\begin{remark}
In the "`quantum"' description one has to differentiate clearly between wave
functions or vectors and the operators or matrices, It is a linear theory,
therefore all operations are simple and do not allow functions of functions.
\end{remark}

The following example illustrates this framework.

\begin{example}
Consider an urn containing 100 balls numbered from 1 through to 100. Basis
states of the world correspond to situations where a ball with a particular
number is drawn from this urn. Hence, the basis states $\Psi _{\omega }$
correspond to states where a ball with a particular number is drawn. General
composite states are complex-valued weighted sums of the basis states $%
|\beta >=\sum_{\omega }\beta _{\omega }\Psi _{\omega }$ and contain
information about the relative frequencies of the outcomes. In this case,
the Hilbert space $H=\mathbb{C}^{100}.$ \newline
Betting 100 Euros on a ball with an even number $\omega $ being drawn from
the urn is an action $A_{\alpha ^{e}}=\underset{\omega }{\sum }\alpha
_{\omega }^{e}P_{\omega }\ $with 
\begin{equation*}
\alpha _{\omega }^{e}=\left\{ 
\begin{tabular}{lll}
$100$ & for & $\omega $ even \\ 
$0$ & for & $\omega $ odd%
\end{tabular}%
\ \right. .
\end{equation*}%
The action $a^{e}$ of betting on an even number induces a utility vector%
\footnote{%
In this example, we consider the special case of a linear utility payoff
function, i.e., of risk neutrality. It would be straightforward to assume a
non-linear utility of payoffs.} over outcomes yielding in state $|\beta >$
the expected utility%
\begin{equation}
U(\alpha ^{e},\beta ):=<\beta |A_{\alpha ^{e}}|\beta >=\underset{%
\begin{tabular}{l}
$\omega \in \Omega $ \\ 
$\omega $ even%
\end{tabular}%
}{\sum }100|\beta _{\omega }|^{2}.
\end{equation}

Similarly, a bet on the odd numbers $P_{\alpha ^{o}}=\underset{\omega }{\sum 
}\alpha _{\omega }^{o}P_{\omega }$ with $\alpha _{\omega }^{o}=\left\{ 
\begin{tabular}{lll}
$100$ & for & $\omega $ odd \\ 
$0$ & for & $\omega $ even%
\end{tabular}%
\right. $ yields an expected utility of $U(a^{o},\beta )=<\beta |A_{\alpha
^{o}}|\beta >$. We assume that, in a state $\beta ,$ a decision maker
chooses among actions $\alpha \in \mathcal{A}:=\{\alpha ^{e},\alpha ^{o}\}$
according to whether $U(\alpha ^{e},\beta )\lessgtr U(\alpha ^{o},\beta ).$
\end{example}

\strut

The following table summarizes the basic framework of a decision problem:

\begin{equation*}
\begin{tabular}{llll}
$\bullet $ & \emph{Hilbert space} & $H:=\mathbb{C}^{|\Omega |},$ &  \\ 
$\bullet $ & \emph{Basis states } & $\Psi _{\omega }:=|\omega >\in H,$ &  \\ 
& (complete set of normalized wave functions ) &  &  \\ 
$\bullet $ & \emph{Composite states of the world } & $|\beta >:=\sum_{\omega
}\beta _{\omega }|\omega >\in H,$ &  \\ 
&(also normalized)
&  &    \\ 
$\bullet $ & \emph{Probability of an outcome }$\omega $ in $\beta $: & $%
<\beta |P_{\omega }|\beta >=|\beta _{\omega }|^{2},$ &  \\ 
&  &    \\ 
$\bullet $ & \emph{Actions} & $A_{\alpha }:=\underset{\omega }{\sum }\alpha
_{\omega }P_{\omega }=\underset{\omega }{\sum }\alpha _{\omega }|\omega
><\omega |,$ &  \\ 
&  &  &  \\ 
$\bullet $ & \emph{Payoffs of action }$\alpha $ & $|\alpha >=A_{\alpha
}|\beta >=\underset{\omega }{\sum }\alpha _{\omega }P_{\omega }|\beta >\in
H, $ &  \\ 
&  &  &  \\ 
$\bullet $ & \emph{Expected payoff of action }$\alpha $ \emph{in state} $%
\beta $: & $U(\alpha ,\beta ):=<\beta |A_{\alpha }|\beta >.$ &  \\ 
&  &  & 
\end{tabular}%
\end{equation*}

The expected payoff of an action $\alpha \ \ $in a composite state $\beta ,\
U(\alpha ,\beta ),$ can be expanded as follows:

\begin{eqnarray*}
U(\alpha ,\beta ) &:&=<\beta |A_{\alpha }|\beta >=<\beta |\underset{\omega }{%
\sum }\alpha _{\omega }P_{\omega }|\beta > \\
&=&\underset{\omega }{\sum }\alpha _{\omega }<\beta |P_{\omega }|\beta >=%
\underset{\omega }{\sum }\alpha _{\omega }|\beta _{\omega }|^{2}.
\end{eqnarray*}

In this framework, an action $\alpha $ assigns payoffs (in terms of utility)
to the outcomes of a composite state.

\strut

As in \cite{Anscombe-Aumann-1963} a state will be associated with a
probability distribution over outcomes. In contrast to \cite%
{Anscombe-Aumann-1963}, however, we do not assume that states can be
observed directly. Instead, the possible states of the world determine the
probabilities over outcomes by projection operators whose expectation values
give the probabilities of the outcomes. A decision maker's information about
possible probability distributions over outcomes is encoded in the states of
the world.

The following example illustrates these concepts in the context of a simple
portfolio choice problem.

\begin{example}[Portfolio choice]
Consider two assets, a stock $a$ and a bond $b.$ The stock pays a state
contingent payoff $(r_{1}q_{0},r_{2}q_{0}).$ The bond pays off the same $r$
in each state. 
\begin{equation*}
\begin{tabular}{l|c|c|c|c|}
& \emph{price} & \emph{quantity} & \multicolumn{2}{|c|}{\emph{payoff}} \\ 
\cline{3-5}
&  &  & state $1$ & state $2$ \\ \hline
\emph{stock:} & $q_{0}$ & $a$ & $r_{1}q_{0}$ & $r_{2}q_{0}$ \\ \hline
\emph{bond:} & $1$ & $b$ & $r$ & $r$ \\ \hline
\end{tabular}%
\end{equation*}%
Basis states are $\Omega =\{1,2\}$ reflecting the per unit return in each
state, $r_{1}>r_{2}.$ The Hilbert space $H$ spanned by the two basis states $%
|1>$ and $|2>$ contains composite states $|\beta >=\beta _{1}|1>+\beta
_{2}|2>$ which capture all information available in state $\beta .$ An
investor chooses a portfolio $\alpha =(a,b)$ subject to a budget constraint $%
q_{0}a+b=W_{0},$ where $W_{0}$ denotes the initial wealth of the investor.
For an investment $\alpha =(a,b)$ and a state $\beta ,$ the investor expects
an expected utility of 
\begin{eqnarray*}
U(\alpha ,\beta ) &=&U(\alpha ,1)|\beta _{1}|^{2}+U(\alpha ,2)|\beta
_{2}|^{2} \\
&=&u(r_{1}a+rb)|\beta _{1}|^{2}+u(r_{2}a+rb)|\beta _{2}|^{2}.
\end{eqnarray*}
\end{example}

\strut Given state $\beta ,$ a decision maker will choose an action $\alpha
^{\ast } $ from a set $\mathcal{A}$ which yields the highest expected
utility $U(\alpha ,\beta ),$ i.e., $U(\alpha ^{\ast },\beta )\geq U(\alpha
,\beta )$ for all $\alpha \in \mathcal{A}.$ Notice, however, that, in
contrast to classical economic decision theory, the optimal choice $\alpha
^{\ast }$ will depend on $\beta $ the state which determines the expected
utility.

\begin{remark}
Since states $\beta $ in the Hilbert space $H$ cannot be observed directly,
observing the action chosen by a decision maker will in general not reveal
the state $\beta $ completely.
\end{remark}

\subsection{Subjective states of mind}

The main conceptual contribution of our paper is the assumption that a
decision maker's subjective attitudes towards the available information can
be modelled by a \emph{subjective state of mind. }States of mind are
elements of the Hilbert space $H$ which represent subjective information
about the environment and the decision makers attitude towards such
information. In particular, the subjective state of mind interacts with the
actual and potential states of the world\footnote{\cite[p. 290]%
{Yukalov-Sornette-2011} introduce the concept of a "space of mind" as the
direct product of "mode spaces, which can be thought of as a possible
mathematical representation of the mind" in order to model mental aspects of
a decision maker. This construct appears to be quite different from our
notion of a mind state.}. In analogy to to a subjective prior in Bayesian
decision making, the subjective state of mind weighs the information
contained in the states and assesses their relevance for the choice of
actions. Other than a subjective Bayesian prior distribution, however, the
notion of a state of mind captures also psychological features of
information processing and distortion. A state of mind represents the
decision maker's perception of the uncertainty. As we will explain in detail
below, the decision maker's perception of uncertainty will change in the
light of new information. Such information is also encoded in elements of
the Hilbert space and will interact with the subjective state of mind. This
interaction allows us to use the power of quantum mechanics to model \emph{%
"entanglements"} and \emph{"superpositions"}.

Formally, a state of mind $\Psi _{M}$ is a composite state $|M>~$in $H.$ It
is associated with the projector $P_{M}=|M><M|$ which projects any state $%
\Psi _{\beta }$ on to $\Psi _{M}$:%
\begin{equation}
P_{M}|\Psi _{\beta }>=c_{\beta }|M>.
\end{equation}

Suppose a decision maker with a state of mind $\Psi _{M}$ is confronted with
information of another state $\Psi _{\beta }.$ Although $\Psi _{\beta }$
represents a normalized state, the moderated state of mind $P_{M}|\Psi
_{\beta }>$ is no longer normalized. In general the coefficient $c_{\beta }$
is complex, but the absolute magnitude $|c_{\beta }|$ of the projection can
be used as a measure of distance in information of state $|\beta >$ from the
subjective "mind state". The more distant the state $|\beta >$ is from the
mind state $|M>$ the smaller the length of the projection (c.f. Fig.1).

\strut

\begin{figure}[h]
\centering
\includegraphics[width=1\textwidth, angle=0]{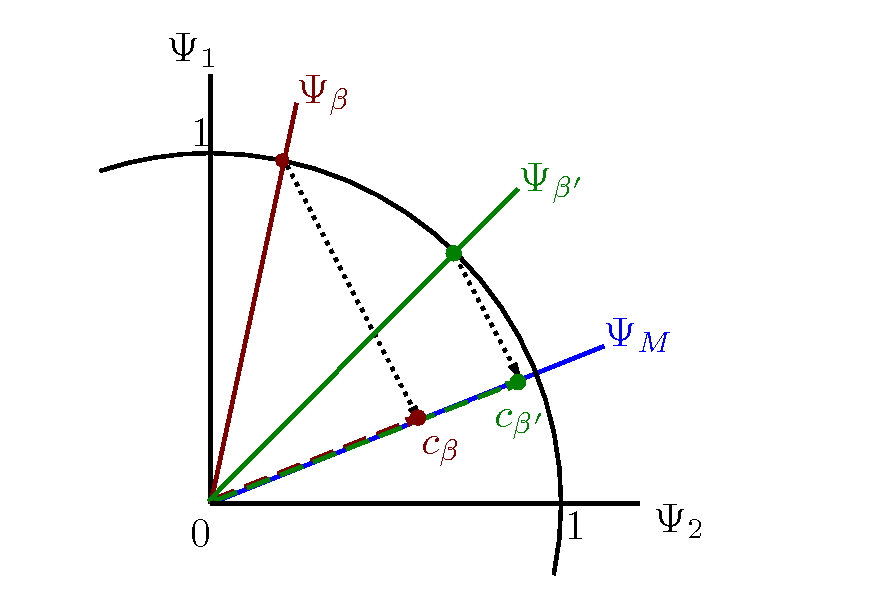}  
\caption{Schematic picture showing the effect of the projection of a composite state on the mind state.
The length of the projection indicates its similarity with the mind state (in general, the coefficients $c$ will be complex
numbers)}
\label{fig-two-states}
\end{figure}

\strut
To assess the information about the basis states $|i> $ contained in an
arbitrary state of the world when the mind state is present, one evaluates
the expected value of the product of the projectors $P_{M}P_{i}P_{M}$. For a
composite state $|\Psi _{\beta }>$ one gets:

\begin{eqnarray}
<\beta |P_{M}P_{i}P_{M}|\beta >\ = &<&\beta |M><M|P_{i}|M><M|\beta >\  \\
&=&|c_{\beta }|^{2}<M|P_{i}|M>
\end{eqnarray}

where%
\begin{equation}
c_{\beta }=<M|\beta >.
\end{equation}

As a result one obtains the probability for the basis state $|i>$ in the
mind state $\Psi _{M}$ multiplied with the absolute square of the amplitude $%
|c_{\beta }|$. The weight $|c_{\beta }|^{2}$ indicates the extent to which
the state $|\beta >$ agrees with the mind state $|M>$. For $|\beta >=|M>$
the decision maker's state of mind will be consistent with the actual state
information, $|c_{\beta }|^{2}=1$.

This framework allows us to study how the subjective state of mind $\Psi _{M}
$ interacts with actual information. The projections on to the basis states
can be interpreted as a core belief and the possible distances $|c_{\beta }|$
as measures of ambiguity regarding the information contained in other states.

\strut

In the following section we will illustrate how this general framework can
model behavior in the two-state Ellsberg paradox.

\section{The Ellsberg experiment}

In a seminal paper published in 1961, \cite{Ellsberg-1961} suggested the
following thought experiment. Consider two urns each containing 100 balls (see Fig.2).
The balls are either black or white. It is known that the first urn contains
exactly 50 black and 50 white balls. For the second urn however, the
composition of the colors is unknown.

\strut

\begin{figure}[h]
\centering
\includegraphics[width=0.66\textwidth, angle=0]{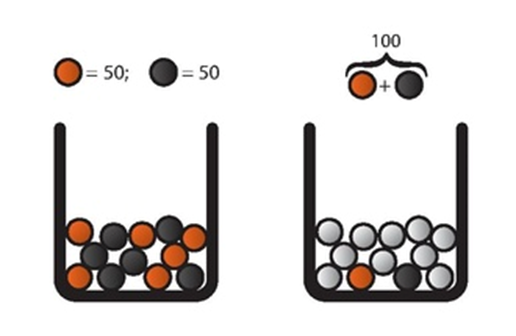}  
\caption{The two Ellsberg urns: Urn 1 on the left contains well defined proportions of black and white balls, whereas for urn 2 on the right the proportions are unknown. }
\label{fig-urns}
\end{figure}

\strut

Subjects hold bets on the color of the balls to be drawn from the urns. A 
\emph{bet on black }$"b"$ yields $100$ Euro if the ball drawn from the urn
is black. Otherwise the payoff is zero. Similarly, \emph{betting on white }$%
"w"$, the subject earns $100$ Euro if a white ball is drawn and nothing
otherwise. Table 1 summarizes this choice problem.

\begin{equation*}
\begin{tabular}{|l|c|c|c|c|}
\hline
& \multicolumn{2}{|c}{\emph{Urn 1}} & \multicolumn{2}{|c|}{\emph{Urn 2}} \\ 
\hline
& 50 & 50 & \multicolumn{2}{|c|}{100} \\ \hline
& \emph{Black} & \emph{White} & \emph{Black} & \emph{White} \\ \hline
$b$ (\emph{bet on black}) & $100$ & $0$ & $100$ & $0$ \\ \hline
$w$ (\emph{bet on white}) & $0$ & $100$ & $0$ & $100$ \\ \hline
\end{tabular}%
\end{equation*}%
Holding a bet first on\emph{\ black} ($b$) and then on \emph{white }($w$),
subjects had to choose the urn on which they wanted to bet.

A large number of subjects, about 60 percent, choose to bet on the draw from
Urn 1 for both bets\footnote{%
The fact that subjects prefer to bet on the urn with the known proportion of
colors could be confirmed in many repetitions of the Ellsberg experiment. 
\cite{Oechssler-Roomets-2015} reviewed 39 experimental studies of the
Ellsberg experiments. They report a median number of 59 percent of ambiguity
averse subjects across all studies which they reviewed.}. These choices
clearly contradict the assumption that these subjects were maximizing SEU.

Denoting by $p_{u}(B)$ the probability of the event $B$ that a black ball is
drawn from urn $u$, $u=1,2,$ the expected utilities of a bet on color $B$
from Urn $u$ is $U_{u}(b)=p_{u}(B)u(100)+\left[ 1-p_{u}(B)\right] u(0)$ and
a bet on $W$ from Urn $u$ is $U_{u}(w)=p_{u}(B)u(0)+\left[ 1-p_{u}(B)\right]
u(100).\ $Obviously, choosing Urn 1 for both bets leads to a contradiction
since 
\begin{equation*}
\begin{tabular}{|l|c|c|c|}
\hline
$\bullet $ & $U_{1}(b)>U_{2}(b)$ & implies & $p_{1}(B)>p_{2}(B),$ \\ \hline
$\bullet $ & $U_{1}(w)>U_{2}(w)$ & implies & $p_{1}(B)<p_{2}(B).$ \\ \hline
\end{tabular}%
\end{equation*}

Thus, the assumption of the decision maker maximizing SEU fails. Notice that
for this conclusion it does not matter what utilities $u(100)$ and $u(0)$
are attached to outcomes, i.e., independent of risk attitude. We only
require $u(100)>u(0)$.

\subsection{Modelling the Ellsberg experiment by a Hilbert space}

The special feature of the Ellsberg experiment, where subjects face the same
bets on the same type of urn, lies in the fact that the information about
the composition of the urn is \emph{precise} in case of Urn 1 and \emph{%
imprecise} in case of Urn 2. We interpret this experimental arrangement in
such a way that the precision of information is reflected in two mind states
of the decision maker, i.e., the two urns exist separately in the
imagination of the experimenter. We therefore model them as two mind states
in the same Hilbert space.

The basis states $|\omega>$ of the Hilbert space consist of the two
exclusive cases (outcomes of the experiment) $\Psi _{W}=|W>,\ i.e. $ \emph{%
"a white ball is drawn from the urn"} and $\Psi _{B}=|B>\ $i.e.,$\ $\emph{"a
black ball is drawn from the urn" }. These basis states are exclusive i.e.
orthogonal and normalized, i.e., $<B|W>=<W|B>=0\ $and $<B|B>=<W|W>=1$.
Composite states $|\beta >=\beta _{B}|B>+\beta _{W}|W>$ of the Hilbert space 
$H\ $spanned by these basis states are characterized by the complex weights $%
\beta =(\beta _{B},\beta _{W}).$

The information of the decision maker regarding the composition of the urns
is encoded in the respective composite states (wave functions\footnote{\cite%
{Giffiths-2002} (p. 203) calls these wave functions "pre-probabilities".})
for the different urns. Urn 1, the urn with the known proportion of $\frac{1%
}{2}$ black and white balls, is described by the composite wave function $%
\Psi _{1}$ 
\begin{equation*}
\Psi _{1}=\frac{1}{\sqrt{2}}\left( \Psi _{B}+\Psi _{W}\right)
\end{equation*}%
and Urn 2, for which the composition is unknown, is described by the wave
function $\Psi _{2}(x)$ 
\begin{equation*}
\Psi _{2}(x)=x\Psi _{B}+\sqrt{1-x^{2}}\Psi _{W}
\end{equation*}

Without reducing the generality of our further arguments we can choose $x$
real, any choice of complex phases in $\Psi _{2}$ can be absorbed in the
final phase of the mind state.

The parameter $x$ contains the information about the composition of Urn 2.
Since it is known that there is a finite number of 100 balls in Urn 2, we
assume that the number of possible states $\Psi _{2}(x)$ of the Hilbert
space $H$ is finite.

The probabilities of the outcomes, a white ball is drawn from the urn $W$ or
a black ball is drawn from the urn $B$, are obtained by the projection
operators $P_{\omega}$, the sum of which yields the identity operator:

\begin{equation*}
\begin{tabular}{lllll}
$P_{B}=|B><B|,$ &  & $P_{W}=|W><W|,$ & with & $P_{B}+P_{W}=1.$%
\end{tabular}%
\end{equation*}%
The probabilities of the outcomes are the expectation values of these
projection operators evaluated with the respective composite state. The
expectation values of the projection operators assign probabilities to the
outcomes given the information of a specific state of the world, $\Psi _{1}$
or $\Psi _{2}(x)$: 
\begin{equation*}
\begin{tabular}{lllll}
$\bullet $ & Urn 1: & $p_{1}(B)=<\Psi _{1}|P_{B}|\Psi _{1}>$ & and & $%
p_{1}(W)=<\Psi _{1}|P_{W}|\Psi _{1}>,$ \\ 
&  &  &  &  \\ 
$\bullet $ & Urn 2: & $p_{2}(B)=<\Psi _{2}(x)|P_{B}|\Psi _{2}(x)>$ & and & $%
p_{2}(W)=<\Psi _{2}(x)|P_{W}|\Psi _{2}(x)>.$%
\end{tabular}%
\end{equation*}

\strut

The two bets are modelled by the operators $A_{B}=u(100)P_{B}+u(0)P_{W}$ and 
$A_{W}=u(0)P_{B}+u(100)P_{W}$ . Hence, given the information about Urn 1
contained in $\Psi _{1},$ one obtains the following expected utilities from
these actions:\ 
\begin{eqnarray*}
U(b,1) &=&<\Psi _{1}|A_{B}|\Psi _{1}>=\frac{1}{2}u(100)+\frac{1}{2}u(0), \\
U(w,1) &=&<\Psi _{1}|A_{W}|\Psi _{1}>=\frac{1}{2}u(0)+\frac{1}{2}u(100).
\end{eqnarray*}%
Similarly, given the information regarding Urn 2, one has 
\begin{eqnarray*}
U(b,x) &=&<\Psi _{2}(x)|A_{B}|\Psi _{2}(x)>=x^{2}u(100)+\left(
1-x^{2}\right) u(0), \\
U(w,x) &=&<\Psi _{2}(x)|A_{W}|\Psi _{2}(x)>=x^{2}u(0)+\left( 1-x^{2}\right)
u(100).
\end{eqnarray*}%
Given this information about the urns and no subjective processing of
information, the Ellsberg puzzle would remain unresolved. A choice of
betting on Urn 1 in both cases would yield for any $x\in \mathcal{R}$, 
\begin{equation*}
\begin{tabular}{|l|c|c|c|}
\hline
$\bullet $ & $U(b,1)>U(b,x)$ & $\Longrightarrow $ & $\frac{1}{2}>x^{2},$ \\ 
\hline
$\bullet $ & $U(w,1)>U(w,x)$ & $\Longrightarrow $ & $\frac{1}{2}<x^{2}.$ \\ 
\hline
\end{tabular}%
\end{equation*}

\subsection{Solution of the paradox: a subjective mind state}

Applying the notion of a mind state  represents the decision makers
assessment of the situation given her information. In case of Urn 2, all the
person knows is the total number of balls in the urn. Hence, her mind state
is a subjectively determined general state from the Hilbert space $H$

\begin{equation*}
\Psi _{M2}(y,d):=y|B>+e^{id}\sqrt{1-y^{2}}|W>
\end{equation*}%
which is characterized by the parameters $(y,d).$ In general, different
decision makers will have different initial states of mind $\Psi _{M2}(y,d).$

All decision makers are however confronted with the same information about
Urn 2 
\begin{equation*}
\Psi _{2}(x)=x\Psi _{B}+\sqrt{1-x^{2}}\Psi _{W}
\end{equation*}%
with the unknown parameter\footnote{%
In case of the information, we will abstract from distortions of the
amplitude.} $x\in \mathcal{R}$.

We assume that the mind state $\Psi _{M2}(y,d)$ distorts the perception of
the basis states of the world represented by the projectors $P_{\omega
}:=|\omega ><\omega |,$ $\left( \omega =B,W\right) $, i.e., the operators to
draw a black or a white ball,

\begin{eqnarray*}
P_{B} &\rightarrow &P_{M2}P_{B}P_{M2}, \\
P_{W} &\rightarrow &P_{M2}P_{W}P_{M2},
\end{eqnarray*}%
where 
\begin{eqnarray*}
P_{M2} &=&|M2><M2| \\
&=&\left[ 
\begin{array}{cc}
y^{2} & e^{-id}y\sqrt{1-y^{2}} \\ 
e^{id}y\sqrt{1-y^{2}} & 1-y^{2}%
\end{array}%
\right]
\end{eqnarray*}%
is the projector of the mind state. For the mind state, the amplitudes
reflect the subjective distortions. Hence, 
\begin{eqnarray*}
\left\langle \Psi _{2}(x)|P_{M2}P_{B}P_{M2}|\Psi _{2}(x)\right\rangle
&=&<\Psi _{2}(x)|M2><M2|P_{B}|M2><M2|\Psi _{2}(x)> \\
&=&y^{2}|c_{x}|^{2} .
\end{eqnarray*}
with 
\begin{eqnarray*}
y^2&=&<M2|P_{B}|M2> \\
|c_{x}|^{2}&=&|<M2|\Psi _{2}(x)>|^2 \\
&=&\left(1-x^{2}-y^{2}+2x^{2}y^{2}+2xy\sqrt{(1-x^{2})(1-y^{2})}\cos {d}%
\right)
\end{eqnarray*}

One sees how the mind state overwrites the state $\Psi _{2}(x)$ of Urn 2,
and modifies the resulting probability with the overlap probability of the
two states. This overlap probability contains an interference term which is
visible from the trigonometric function. It encodes the reliability the
person assigns to his/her estimate of this probability. Similarly, one
obtains the probability of $W$ from the distorted state as

\begin{eqnarray*}
\left\langle \Psi _{2}(x)|P_{M2}P_{W}P_{M2}|\Psi _{2}(x)\right\rangle
&=&<\Psi _{2}(x)|M2><M2|P_{W}|M2><M2|\Psi _{2}(x)> \\
&=&(1-y^{2})|c_{x}|^{2} .
\end{eqnarray*}

with $|c_{x}|^{2}$ as above and

\begin{eqnarray*}
1-y^2&=&<M2|P_{W}|M2>
\end{eqnarray*}

In the case of Urn 1, the decision maker knows the composition of the urn.
Hence, her mind state should be one of subjective certainty 
\begin{equation*}
|\Psi _{M1}>=|\Psi _{1}>=\frac{1}{\sqrt{2}}\left( \Psi _{B}+\Psi _{W}\right)
\end{equation*}%
where the state of mind $|\Psi _{M1}>$ coincides with the actual information 
$|\Psi _{1}>$.

Hence, we have 
\begin{equation*}
<\Psi _{1}|P_{M1}P_{\omega }P_{M1}|\Psi _{1}>=|<\Psi _{M1}|\Psi
_{1}>|^2<\Psi _{M1}|P_{\omega }|\Psi _{M1}>=\frac{1}{2}
\end{equation*}%
for $\omega =W$ and for $\omega =B.$

Given the action projectors of the two bets $A_{B}=u(100)P_{B}+u(0)P_{W}$
and $A_{W}=u(0)P_{B}+u(100)P_{W},$ one obtains as the conditions for
Ellsberg behavior:

\begin{itemize}
\item for the bet on Black $b,$ Urn 1 is preferred if 
\begin{eqnarray*}
U(b,1,M1) >U(b,x,M2)
\end{eqnarray*}

with

\begin{eqnarray*}
U(b,1,M1) &=&\left\langle \Psi _{1}|P_{M1}A_{B}P_{M1}|\Psi _{1}\right\rangle
\\
&=&\frac{1}{2}\left[ u(100)+u(0)\right] , \\
U(b,x,M2) &=&|c_{x}|^{2}\left[ y^{2}u(100)+\left( 1-y^{2}\right) u(0)\right],
\end{eqnarray*}

\item for a bet on White $w,$ Urn 1 is preferred if 
\begin{equation*}
U(w,1,M1)>U(w,x,M2)
\end{equation*}

with

\begin{eqnarray*}
U(w,1,M1) &=&\left\langle \Psi _{1}|P_{M1}A_{W}P_{M1}|\Psi _{1}\right\rangle
\\
&=&\frac{1}{2}\left[ u(100)+u(0)\right] , \\
U(w,x,M2) &=&|c_{x}|^{2}\left[ y^{2}u(0)+\left( 1-y^{2}\right) u(100)\right]
.
\end{eqnarray*}
\end{itemize}

\strut

For $u(0)=0$ and $u(100)=1,$ we obtain the following conditions:%
\begin{eqnarray*}
\frac{1}{2} &>&|c_{x}|^{2}y^{2}, \\
\frac{1}{2} &>&|c_{x}|^{2}\left( 1-y^{2}\right) .
\end{eqnarray*}

Assuming no effect of the interference term $d=\frac{\pi }{2},$ i.e., $\cos {%
d=0,}$ Fig.3 shows the regions for the Ellsberg choices.

\begin{figure}[tbp]
\centering
\includegraphics[width=0.7\linewidth]{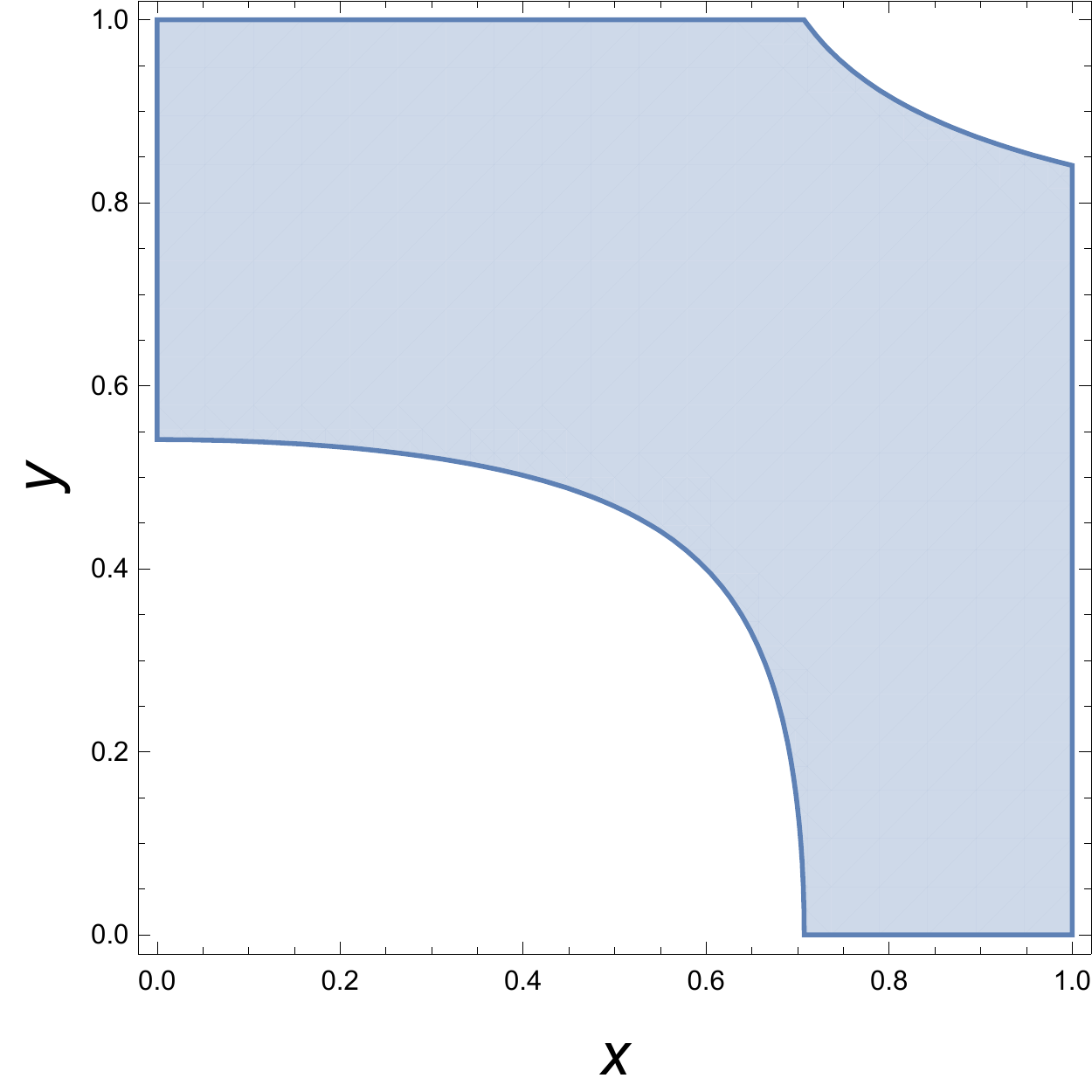}
\caption{The figure shows for fixed $d=\frac{\protect\pi}{2}$ the region $%
\frac{1}{2} >|c_{x}|^{2}y^{2}$ and $\frac{1}{2} >|c_{x}|^{2}\left(
1-y^{2}\right)$ as grey ( in color blue) shaded area.}
\label{fig:spi2.pdfq
}
\end{figure}

On the horizontal axis we consider values of $x\in \lbrack 0,1]$ and on the
vertical axis of $y\in \lbrack 0,1].$ In this diagram we consider the case
without interference due to the complex conjugate since $d=\frac{\pi }{2}\ $%
implies$\ \cos {d=0.}$ In this case, $|c_{x}|^{2}$ simplifies to $\left(
1-x^{2}-y^{2}+2x^{2}y^{2}\right) .$ The grey ( in color blue) area contains $(x,y)$
combinations which correspond to Ellsberg behavior. Integrating, one obtains
an area of approximately 63 percent. In our interpretation, the parameter $y$
captures the decision maker's initial disposition in regard to Urn 2. The
parameter $x,$ which is measured as the unknown actual probability of the
draw of a black ball, captures the objective circumstances of the situation.
Projecting $\left( x,\sqrt{1-x^{2}}\right) $ onto the mind state $\left(
y,e^{\frac{\pi }{2}i}\sqrt{1-y^{2}}\right) $ yields a mind state which is
affected by the actual situation.

In their review of 39 experimental studies of the Ellsberg paradox, \cite%
{Oechssler-Roomets-2015} find a wide range of measured ambiguity across the
various studies. These different experimental contexts which include the
different ways of actually choosing the proportion on the balls in the
unknown urn reflects the kind of objective experimental environment which
meets the individual states of minds of the subjects. The resulting behavior
is influenced by both factors, the subjective mind state $y$ and the
objective situation $x.$ Measurements such as the observed behavior and the
stated predictions regarding the proportions of the colors should reflect
these parameters. This will be discussed in the next section.

\section{Discussion and Conclusion}

The Ellsberg paradox (\cite{Ellsberg-1961}) is usually interpreted as a
manifestation of the decision maker's ambiguity about the unknown
proportions in Urn 2. Faced with a choice between bets on draws from an urn
with a known probability distribution and bets on an urn with an unknown
probability distribution people prefer to bet on draws from the urn with the
known probability distribution. In the two urn problem individual subjects
select for all bets the urn where white and a black balls are equally
distributed. Such behavior is inconsistent with maximization of expected
utility for any subjective probability distribution regarding the
composition of Urn 2. In the literature on decision making under
uncertainty, most of the suggested solutions of the paradox assume
"ambiguity aversion" of the decision maker, i.e., the decision maker
evaluates bets on Urn 2 according to the worst case of all possible
compositions.

The model which we put forward in this paper goes beyond this psychological
explanation by "pessimism" and invokes the framework of a Hilbert space as
possibility space which captures various aspects of the choice situation
including, but not being confined to, information about the composition of
the urns. The Hilbert space contains the two possible draws from the urns, a
black ball or a white ball, as basis states. Based on the information about
the urns, one constructs wave functions for the respective urn states from
these basis states. These wave functions may contain complex amplitudes
entangling the basis states. Hence, there is more information in these wave
functions than in real probabilities.

In applications to a decision-theoretic context, a crucial role is played by
the subjective mind states which reflect the mental perception
(consciousness) of the subjects in the experiment. The precise information
about the content of Urn 1 we represent by a mind state equal to a
preprobability without complex amplitude. For the second urn, however, the
mind state represents other possibilities than the one given by the unknown
objective composition of the second urn. In this case the parameters
describe the influence of beliefs, information, and other context variables
on the state of the second urn. The single complex phase in the mind state
summarizes all possible effects of complex amplitudes for the second-urn
wave functions.

For the three parameters $(x,y,d)$, the condition for choices corresponding
to the typical Ellsberg behavior defines a subspace of the feasible space of
the three parameters. Fig. 4 illustrates the entanglement due
to the complex amplitude for the parameter space of our model $[0,1]\times
\lbrack 0,1]\times \lbrack 0,\pi ]$.

\begin{figure}[tbp]
\centering
\includegraphics[width=0.7\linewidth]{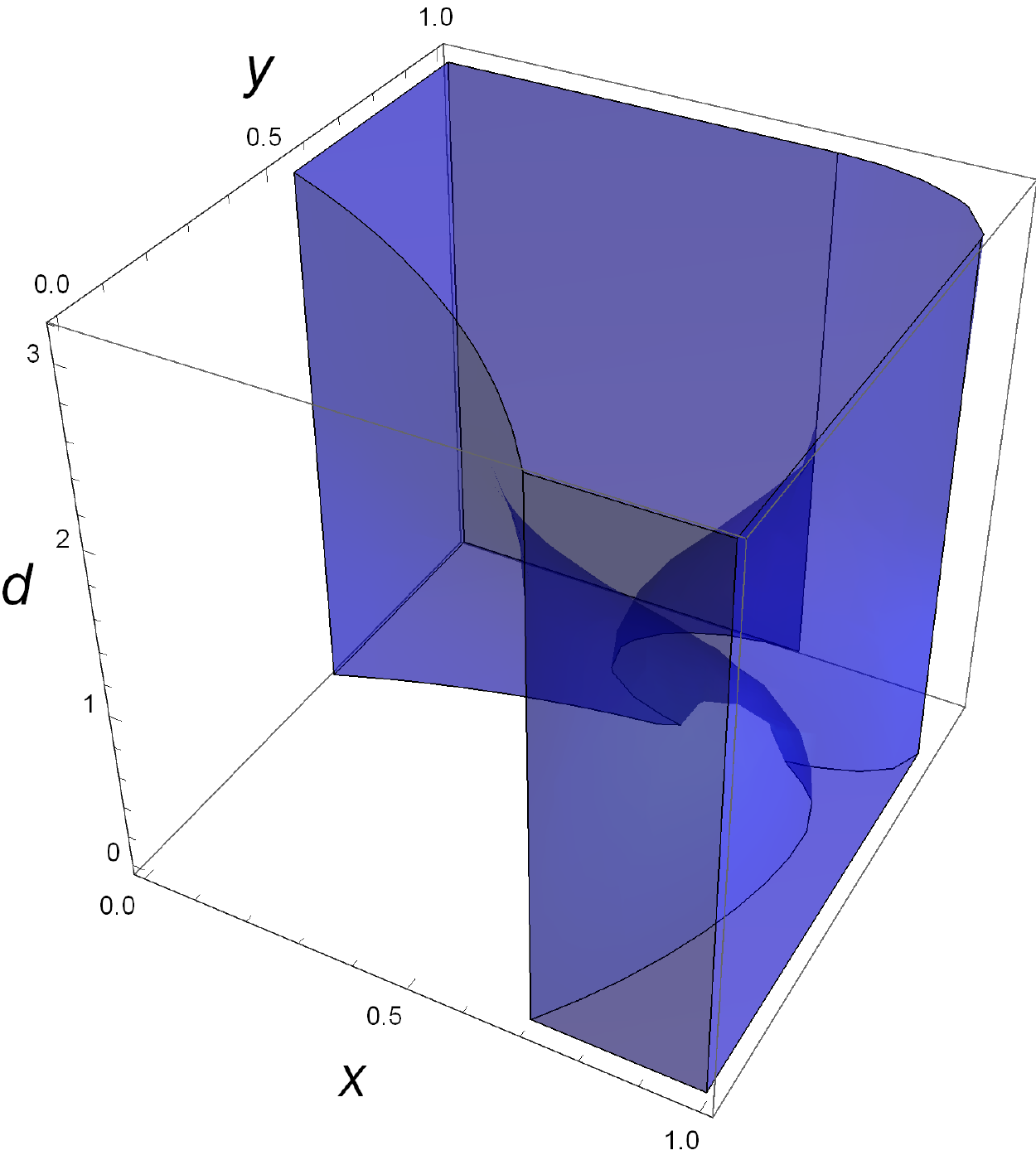}
\caption{The figure shows  the grey (in color blue) region in the three parameter space of 
$x,y,d$ connected to the objective and subjective second urn state with $%
\frac{1}{2} >|c_{x}|^{2}y^{2}$ and $\frac{1}{2} >|c_{x}|^{2}\left(
1-y^{2}\right)$}
\label{fig:u3D.pdf}
\end{figure}

\strut

In a first attempt to summarize the results about ambiguity-averse behavior
across the 39 Ellsberg experiments in their review, \cite%
{Oechssler-Roomets-2015} obtain an average ratio $\ r_{emp}$ of
approximately 57 percent ambiguity averse subjects\footnote{%
The interval for the empirical result does not reflect a statistical error.
It is derived from the average over all existing experiments and over a
restricted list omitting experiments with the highest and lowest results.},

\begin{equation*}
r_{emp}\in \lbrack 57.2\%,57.9\%].
\end{equation*}

Using the ratio of the volume of the subspace of parameters leading to
ambiguity-averse choices to the total volume as a naive measure for the
percentage of ambiguity averse decisions in the Ellsberg two-urn problem
yields the ratio $r_{quant},$%
\begin{equation*}
r_{quant}=58.1\%\pm 0.5\%.
\end{equation*}

Though one obtains a surprisingly similar proportion, we do not claim that
aggregating over all parameter values is an appropriate measure for
ambiguity averse behavior in our model. A more careful analysis may restrict
the range of reasonable mind states and their complex amplitude $(y,d)$.

The Hilbert space method which we suggest in this paper increases the number
of parameters in general. On the one hand, this allows us to explain results
which are paradoxical in classical decision theory, on the other hand, this
exposes the model to the danger of arbitrariness. In especially designed
experiments, however, one could ask more questions in order to study the
role of these additional parameters separately. E.g., in the two urn problem
one could ask the subjects for their estimate of the number of black and
white balls in the second urn. This would allow one to fix one of the
parameters of the wave function of the mind state, thus providing a more
stringent test of the model.

In Economics and most of the Social Sciences, classical decision theory
under uncertainty in the framework of \cite{Savage-1954} has been extremely
fruitful, allowing economists to develop new fields such as information
economics, contract theory, and financial markets analysis. Human choice
behavior as observed in experiments, however, has cast some doubts on the
general accuracy of the Savage approach as a description of actual behavior.
Yet most existing generalizations of the SEU model maintain the basic
framework of actions mapping states into consequences.

The main conceptual difference between the Savage framework and the Hilbert
space approach advanced in quantum mechanics is the notion of a "state". 
\cite{Savage-1954} \ defines "a \emph{state} (of the world)" as "a
description of the world, leaving no relevant aspect undescribed" and "the 
\emph{true} state of the world" as "the state that does in fact obtain,
i.e., the true description of the world" (p. 9). In the Savage context, the
true state is revealed unambiguously, e.g., in the Ellsberg paradox the "
true state of the world" is represented by the color of the ball drawn from
an urn.

In quantum mechanics a "state of the system" cannot be observed directly.
What is observed are measurements. In general, a state of a system contains
more information than can be observed in one measurement. Thus, the Hilbert
space approach takes into account that states of the world may be too
complex to be described or observed in its entirety. From this perspective,
the state of the world in the Ellsberg example comprises the complete
environment of two urns, the composition of these urns, the information
about the urns, the psychological interpretations of the subjects, etc. The
observed "color of the ball drawn from the urn" is a measurement.

Even in the experimental environment of a laboratory, the (Savage) notion of
a state of the world has proved to be quite elusive. Framing and nudging are
well-known phenomena which contradict the assumption that there is a "state
of the world" which can be treated as unrelated to the context of the
experiment. Hence, studying complex Hilbert spaces which generate
entanglements between subjective beliefs, information, and other aspect of
the environment may be a useful exercise. We view the possibility space as a
first step towards such new concepts.

Finally, classical decision theory is based on the maximization of utility
and probability theory which allows for information dynamics according to
the Bayesian updating rule. The Hilbert space approach allows also for a
dynamic evolvement. Quantum mechanics has been successful because Erwin Schr%
\"{o}dinger introduced a dynamics in the Hilbert space which determines the
evolution of states over time. In neurophysiology, we are far from
understanding this dynamics. Exploring an Hilbert space of states may
present a first step towards understanding the evolvement of the human
perception of a complex environment.

\strut

\bibliographystyle{econometrica}
\ifx\undefined\BySame
\newcommand{\BySame}{\leavevmode\rule[.5ex]{3em}{.5pt}\ }
\fi
\ifx\undefined\textsc
\newcommand{\textsc}[1]{{\sc #1}}
\newcommand{\emph}[1]{{\em #1\/}}
\let\tmpsmall\small
\renewcommand{\small}{\tmpsmall\sc}
\fi

\strut

\section*{Appendix}

\subsection*{A: A short primer on notation and computations}

The basis wave functions in the two dimensional Hilbert space are
represented by Dirac kets $|>$:%
\begin{equation*}
\begin{tabular}{lll}
$\Psi _{B}=|B>$ & and & $\Psi _{W}=|W>.$%
\end{tabular}%
\end{equation*}%
They may be written as two dimensional column vectors:%
\begin{equation*}
\begin{tabular}{lll}
$|B>=\left( 
\begin{array}{c}
1 \\ 
0%
\end{array}%
\right) $ & and & $|W>=\left( 
\begin{array}{c}
0 \\ 
1%
\end{array}%
\right) .$%
\end{tabular}%
\end{equation*}

Two normalized wave functions characterize the two urns:%
\begin{equation*}
\Psi _{1}=\left( 
\begin{array}{c}
\frac{1}{\sqrt{2}} \\ 
\frac{1}{\sqrt{2}}%
\end{array}%
\right) ,
\end{equation*}

\begin{equation*}
\Psi _{2}(x)=\left( 
\begin{array}{c}
x \\ 
\sqrt{1-x^{2}}%
\end{array}%
\right) .
\end{equation*}%
The parameter $x$ reflects the unknown number of black and white balls in
Urn 2.

For the mind state, however, complex amplitudes matter. Hence, the "mind
state" $|\Psi _{M}>$ will be represented as a superimposition of the black
and white basis elements. Here we use a complex amplitude. The phases of the
other states can be absorbed in the phase $d$ for the final result:

\begin{equation*}
\Psi _{M}=\left( 
\begin{array}{c}
y \\ 
e^{id}\sqrt{1-y^{2}}%
\end{array}%
\right) .
\end{equation*}

\textbf{Complex conjugation}

The elements of the dual space are obtained by complex conjugation and
transposition. For the black and white states, complex
conjugation does not matter. Hence,%
\begin{equation}
\begin{tabular}{lll}
$<B|=\left( 1,0\right) $ & and & $<W|=\left( 0,1\right) .$%
\end{tabular}
\notag
\end{equation}%
For the mind state, however, one obtains 
\begin{equation}
<\Psi _{M}|=\left( y,e^{-id}\sqrt{1-y^{2}}\right) .  \notag
\end{equation}%
Notice the complex conjugation in the latter case.

\textbf{Projection operators}

Projection operators have the property $P^{2}=P$. They correspond to
observables and will be realized in measurements, like drawing a black ball
or drawing a white ball. The projection operator for drawing a black ball $%
P_{B}=|B><B|$ is given by the outer product (tensor product) of the vector $%
|B>$ of the original Hilbert space with the vector $<B|$ in the dual Hilbert
space, i.e.,%
\begin{equation*}
P_{B}=\left( 
\begin{array}{cc}
1 & 0 \\ 
0 & 0%
\end{array}%
\right) .
\end{equation*}

Similarly, the projection operator $P_{W}$ on white has the form: 
\begin{equation*}
P_{W}=\left( 
\begin{array}{cc}
0 & 0 \\ 
0 & 1%
\end{array}%
\right) .
\end{equation*}

The probability to obtain a black ball in urn 1 can be obtained from the
expectation value of the projection operator on Black with the wave function
of Urn 1:%
\begin{eqnarray*}
p_{1}(B) &=&<\Psi _{1}|P_{B}|\Psi _{1}> \\
&=&\left( \frac{1}{\sqrt{2}},\frac{1}{\sqrt{2}}\right) \left( 
\begin{array}{cc}
1 & 0 \\ 
0 & 0%
\end{array}%
\right) \left( 
\begin{array}{c}
\frac{1}{\sqrt{2}} \\ 
\frac{1}{\sqrt{2}}%
\end{array}%
\right) =\frac{1}{2}.
\end{eqnarray*}

It involves the multiplication of three matrices: one row vector of length 2
for $<\Psi _{1}|$, the $2x2$ matrix for the projection operator, and the
column vector of length 2 for $|\Psi _{1}>$. The probability of drawing a
black ball from Urn 2 is computed similarly as 
\begin{eqnarray*}
p_{2}(B) &=&<\Psi _{2}(x)|P_{B}|\Psi _{2}(x)> \\
&=&\left( x,\sqrt{1-x^{2}}\right) \left( 
\begin{array}{cc}
1 & 0 \\ 
0 & 0%
\end{array}%
\right) \left( 
\begin{array}{c}
x \\ 
\sqrt{1-x^{2}}%
\end{array}%
\right) =x^{2}.
\end{eqnarray*}

Replacing the matrix $P_{B}$ by the matrix $P_{W},$ one obtains the
respective expressions for the probabilities of drawing a white ball from
the urns. In general, the state of a system can only be described by such
measurements.

The mind state projects the state of the urns onto the mind state by the
projection operator $P_{M}$,%
\begin{eqnarray*}
P_{M} &=&|\Psi _{M}><\Psi _{M}| \\
&=&\left( 
\begin{array}{cc}
y^{2} & e^{-id}y\sqrt{1-y^{2}} \\ 
e^{id}y\sqrt{1-y^{2}} & 1-y^{2}%
\end{array}%
\right) .
\end{eqnarray*}%
The subjective probability of drawing a white ball from Urn 2 given the mind
state $\Psi _{M}$ is obtained as%
\begin{equation*}
\begin{tabular}{lll}
& $\left\langle \Psi _{2}(x)|P_{M}P_{W}P_{M}|\Psi _{2}(x)\right\rangle $ & 
\\ 
&  &  \\ 
& $=<\Psi _{2}(x)|\Psi _{M}><\Psi _{M}|P_{W}|\Psi _{M}><\Psi _{M}|\Psi
_{2}(x)>$ &  \\ 
&  &  \\ 
& 
\fontsize{7pt}{5}\selectfont%
$=\underset{<\Psi _{2}(x)|\Psi _{M}>}{\underbrace{\left[ \left( x,\sqrt{%
1-x^{2}}\right) \left( 
\begin{array}{c}
y \\ 
e^{id}\sqrt{1-y^{2}}%
\end{array}%
\right) \right] }}\underset{<\Psi _{M}|P_{W}|\Psi _{M}>}{\underbrace{\left[
\left( y,e^{-id}\sqrt{1-y^{2}}\right) \left( 
\begin{array}{cc}
0 & 0 \\ 
0 & 1%
\end{array}%
\right) \left( 
\begin{array}{c}
y \\ 
e^{id}\sqrt{1-y^{2}}%
\end{array}%
\right) \right] }}\underset{<\Psi _{M}|\Psi _{2}(x)>}{\underbrace{\left[
\left( y,e^{-id}\sqrt{1-y^{2}}\right) \left( 
\begin{array}{c}
x \\ 
\sqrt{1-x^{2}}%
\end{array}%
\right) \right] }}$ &  \\ 
&  &  \\ 
& 
\fontsize{11pt}{5}\selectfont%
$=\left( 1-x^{2}-y^{2}+2x^{2}y^{2}+2xy\sqrt{(1-x^{2})(1-y^{2})}\cos {d}%
\right) (1-y^{2}).$ & 
\end{tabular}%
\end{equation*}

All expressions in the main paper are computable by these matrix operations.

\newpage

\subsection*{B: Diagrams}

The following diagrams show the parameter regions of Ellsberg behavior, i.e.
the overlap of the regions which represent 
\begin{equation*}
\begin{tabular}{lll}
$\bullet $ & the preference for betting on black $(b)$ in Urn 1: & $\frac{1}{%
2}>|c_{x}|^{2}y^{2},$ \\ 
& and &  \\ 
$\bullet $ & the preference for betting on white $(w)$ in Urn 1: & $\frac{1}{%
2}>|c_{x}|^{2}\left( 1-y^{2}\right) ,$%
\end{tabular}%
\end{equation*}%
for different values of the parameter $d.$ Below the diagrams, we give the
ratios of the grey (in color blue) areas relative to the total areas in percentages.

\strut 
\begin{equation*}
\begin{tabular}{ccc}
$d=\frac{0}{4}\ \pi $ &  & $d=\frac{1}{4}\ \pi $ \\ 
\begin{minipage}{.3\textwidth} \includegraphics[width=\linewidth,
height=30mm]{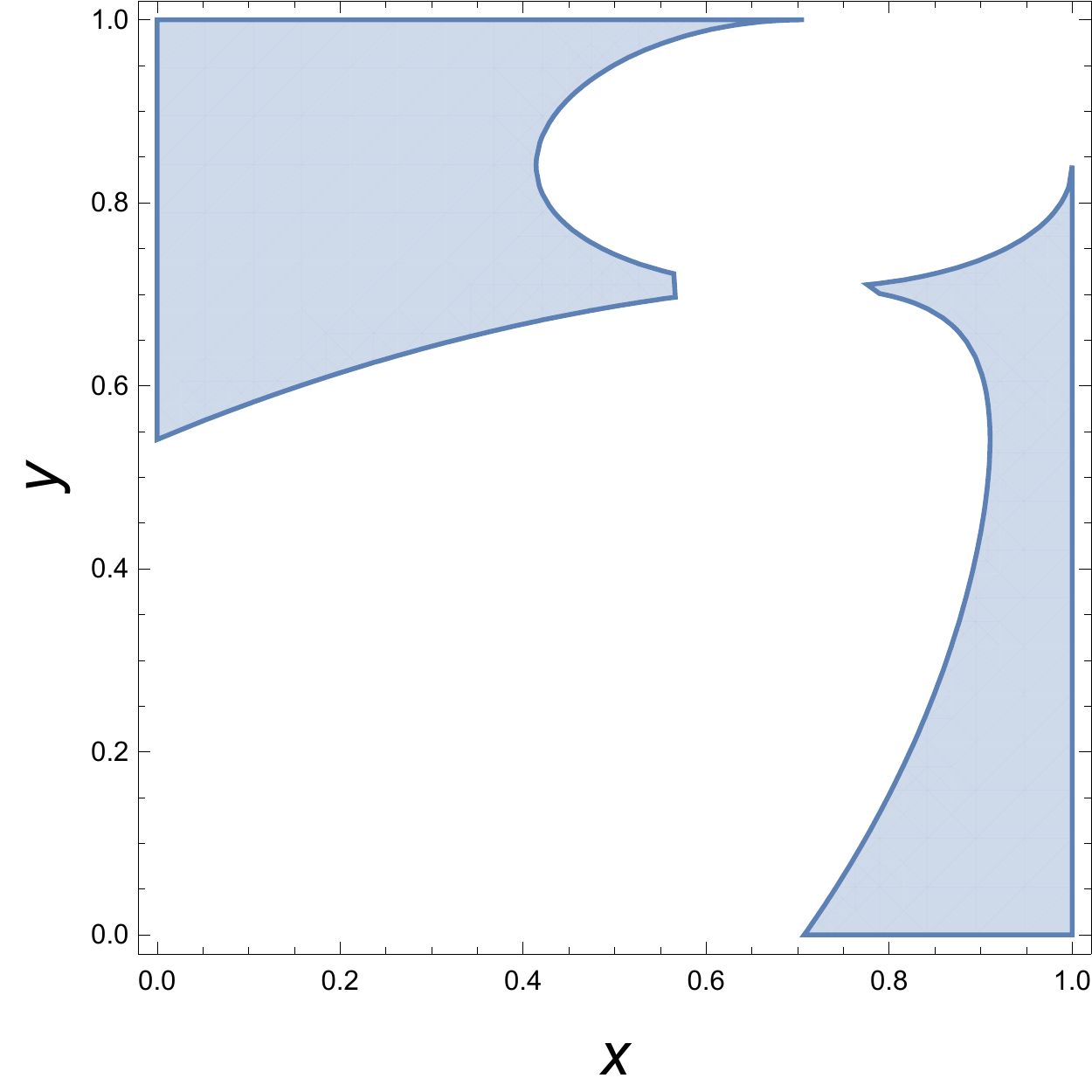} \end{minipage} &  & \begin{minipage}{.3\textwidth}
\includegraphics[width=\linewidth, height=30mm]{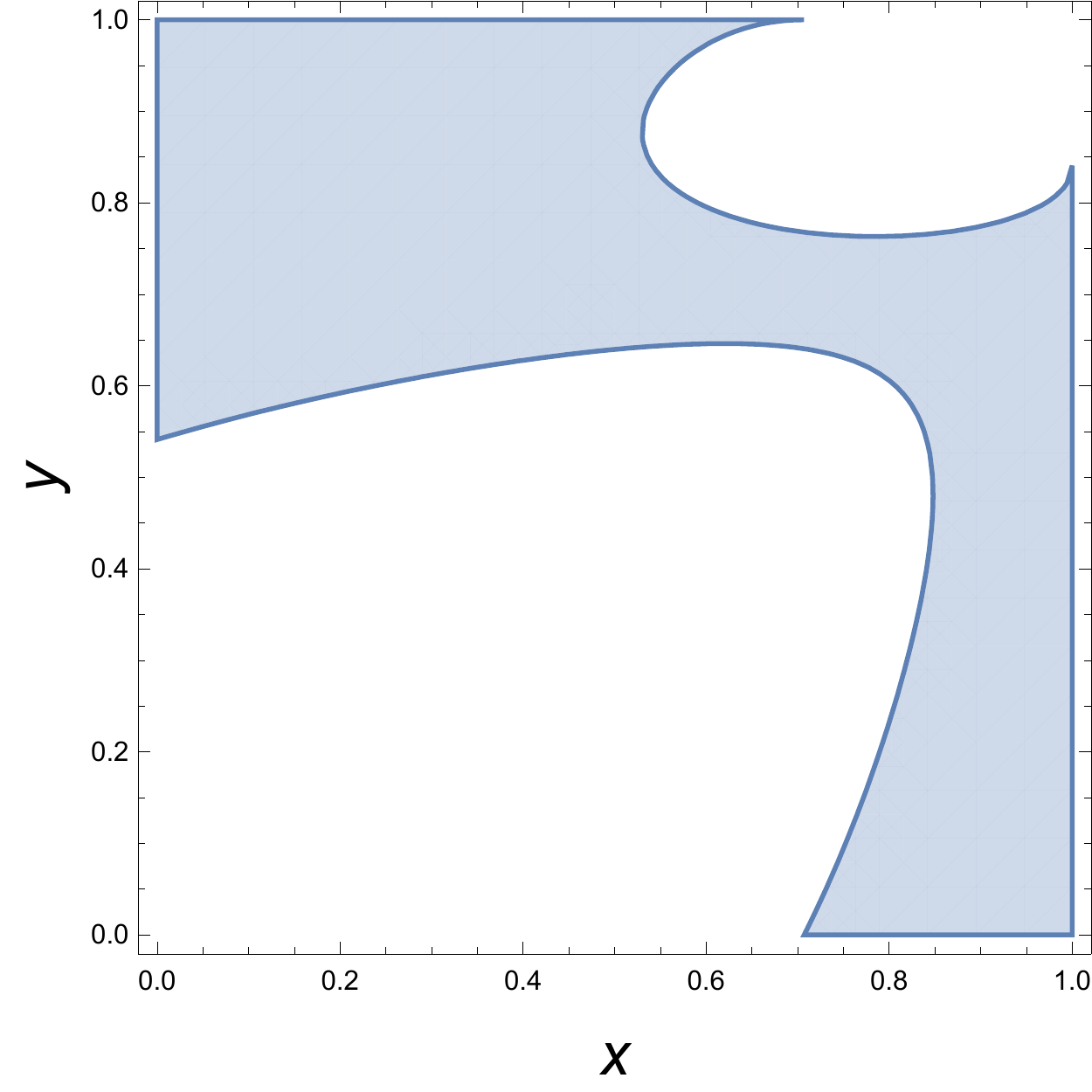} \end{minipage} \\ 
Grey (in color blue) area: $30 \%$ &  & Grey (in color blue) area : $41 \%$ \\ 
& $d=\frac{2}{4}\pi $ &  \\ 
& \begin{minipage}{.3\textwidth} \includegraphics[width=\linewidth,
height=30mm]{spi2.pdf} \end{minipage} &  \\ 
& Grey (in color blue) area: $63 \%$ &  \\ 
$d=\frac{3}{4}\pi $ &  & $d=\frac{4}{4}\pi $ \\ 
\begin{minipage}{.3\textwidth} \includegraphics[width=\linewidth,
height=30mm]{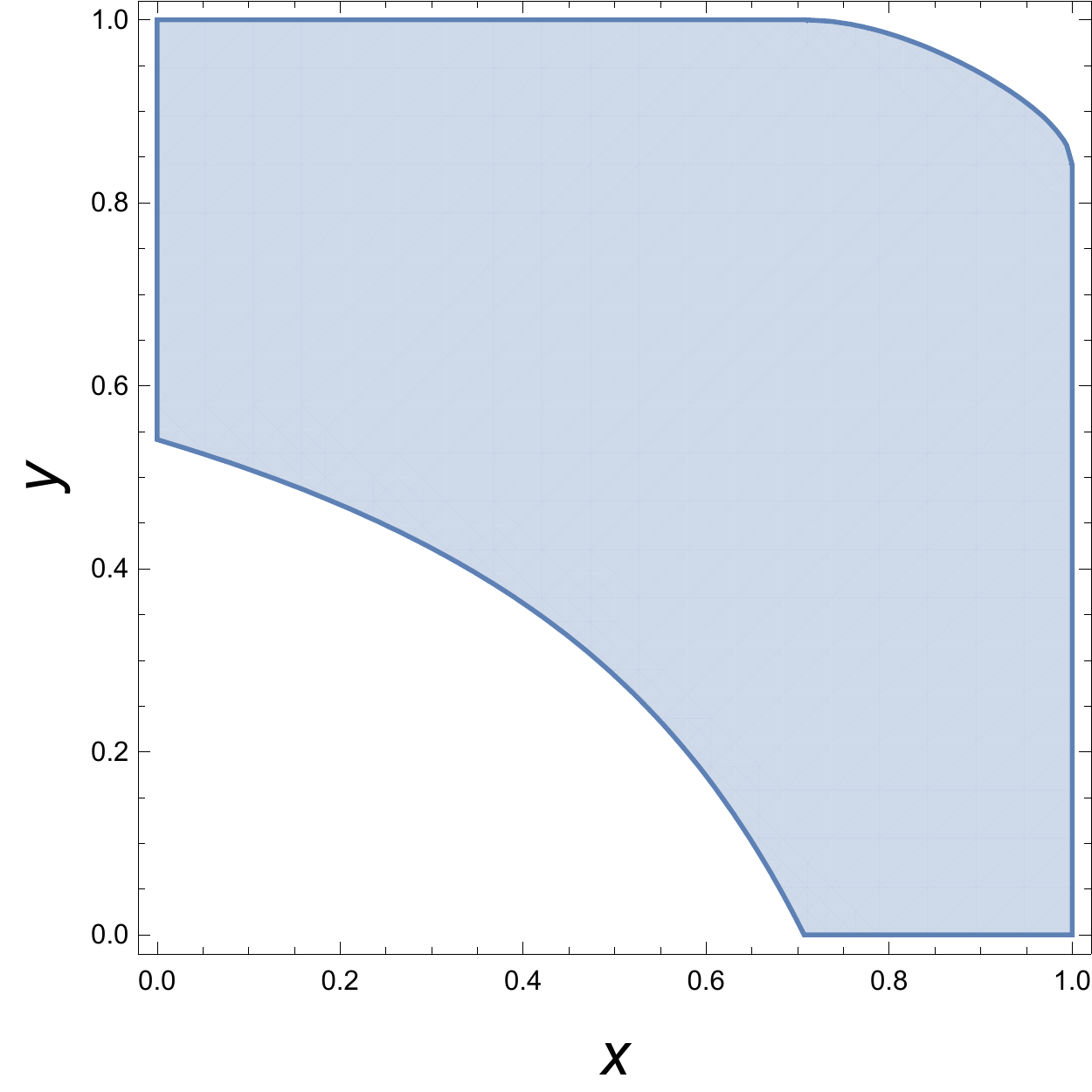} \end{minipage} &  & \begin{minipage}{.3\textwidth}
\includegraphics[width=\linewidth, height=30mm]{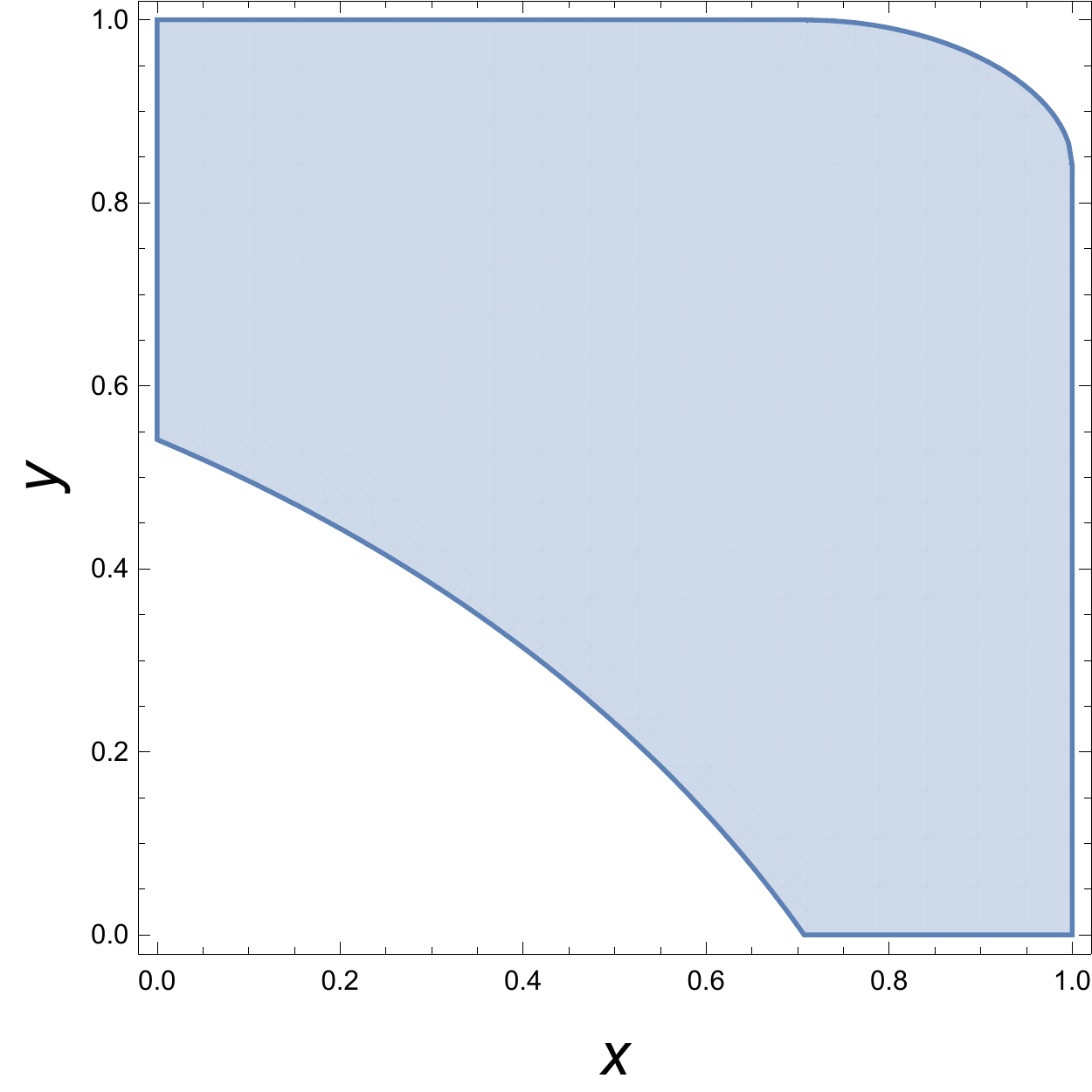} \end{minipage} \\ 
Grey (in color blue) area: $74 \%$ &  & Grey (in color blue) area: $76 \%$%
\end{tabular}%
\end{equation*}

\strut

\newpage

\strut

\textbf{Highlights:}

\begin{itemize}
\item The Hilbert space method used in quantum theory is applied to decision
making under uncertainty.

\item States of the world may be too complex to be described or observed in
its entirety.

\item  The potential of the approach to deal with well-known paradoxa of
decision theory is demonstrated in the context of the Ellsberg two-urn
paradox.
\end{itemize}

\strut

\end{document}